\documentclass[a4paper, UKenglish, cleveref, autoref, thm-restate]{lipics-v2021}

\usepackage{graphicx}
\usepackage{amsmath}
\usepackage{amssymb}
\usepackage{amsthm}
\usepackage{stmaryrd}
\usepackage[dvipsnames]{xcolor}
\usepackage{hyperref}
\usepackage{xspace}
\usepackage{comment}
\usepackage{mathrsfs}
\usepackage{multirow}

\theoremstyle{plain}

\theoremstyle{definition}

\newcommand{\id}{\mathrm{id}}

\newcommand{\mat}{\mathrm{M}}
\newcommand{\gl}{\mathrm{GL}}
\newcommand{\N}{\mathbb{N}}
\newcommand{\Z}{\mathbb{Z}}
\newcommand{\Q}{\mathbb{Q}}
\newcommand{\set}[1]{\left\{#1\right\}}

\newcommand{\tr}{\mathrm{tr}}

\newcommand{\spn}{\mathrm{Span}}
\newcommand{\enve}{\mathrm{Env}}

\DeclareMathOperator{\BlockDiag}{BlockDiag}

\DeclareMathOperator{\bitsize}{\mathrm{size}}

\DeclareMathOperator{\Imag}{im}
\DeclareMathOperator{\rk}{rk}

\newcommand{\tmap}{\mathrm{Tr}}
\newcommand{\tmatrix}{T}
\newcommand{\tnf}{Wedderburn-Malcev normal form\xspace}

\newcommand{\magn}[1]{\llbracket #1 \rrbracket}

\newcommand{\tmagn}{height\xspace}
\newcommand{\Tmagn}{Height\xspace}

\newcommand{\cof}{\mathrm{cof}}

\usepackage{tikz}
\usetikzlibrary{shapes.geometric}
\usetikzlibrary{decorations.pathmorphing,decorations.pathreplacing,decorations.shapes}

\newcommand{\ptime}{\textsc{PTime}\xspace}
\newcommand{\np}{\textsc{NP}\xspace}

\newcommand{\pspace}{\textsc{PSpace}\xspace}

\newcommand{\conexp}{\textsc{coNExp}\xspace}
\newcommand{\nexp}{\textsc{NExp}\xspace}

\newcommand{\ie}{\textit{i.e.~}}

\title{Revisiting Finiteness of Matrix Monoids}

\author{Rida Ait El Manssour}{University of Oxford, United Kingdom}{rida.aitelmanssour@cs.ox.ac.uk}{https://orcid.org/0000-0001-6228-9071}{Supported by the EPSRC Fellowship (EP/X033813/1).}

\author{Roland Guttenberg}{University of Warsaw, Poland}{r.guttenberg@uw.edu.pl}{https://orcid.org/0000-0001-6140-6707}{Supported by the ERC grant INFSYS, agreement No. 950398.}

\author{Nathan Lhote}{Aix-Marseille Universit\'{e}, LIS, France}{nathan.lhote@lis-lab.fr}{https://orcid.org/0000-0003-3303-5368}{}

\author{Mahsa Shirmohammadi}{CNRS \& IRIF, France}{mahsa@irif.fr}{
https://orcid.org/0000-0002-7779-2339}{Supported by the ERC Synergy Grant VePaSS (agreement No. 101224640) and the grant VeSyAM (ANR-22-CE48-0005).}

\author{James Ben Worrell}{University of Oxford, United Kingdom}{james.worrell@cs.ox.ac.uk}{https://orcid.org/0000-0001-8151-2443}{Supported by the EPSRC Fellowship (EP/X033813/1).}

\authorrunning{R. A. E. Manssour, R. Guttenberg, N. Lhote, M. Shirmohammadi, J.B. Worrell} 

\Copyright{Rida Ait El Manssour, Roland Guttenberg, Nathan Lhote, Mahsa Shirmohammadi, James Ben Worrell} 



\category{Track B: Automata, Logic, Semantics, and Theory of Programming}

\nolinenumbers 

\EventEditors{Sayan Bhattacharya, Danupon Nanongkai, Michael Benedikt, and Gabriele Puppis}
\EventNoEds{4}
\EventLongTitle{53rd International Colloquium on Automata, Languages, and Programming (ICALP 2026)}
\EventShortTitle{ICALP 2026}
\EventAcronym{ICALP}
\EventYear{2026}
\EventDate{July 7--10, 2026}
\EventLocation{Royal Holloway, University of London, Egham, United Kingdom}
\EventLogo{}
\SeriesVolume{374}
\ArticleNo{170}

\date{}

\begin{document}

\maketitle
\begin{abstract}
This paper concerns decision problems related to finite monoids of rational matrices. 
We show that determining finiteness of a given finitely presented monoid is in \pspace, improving the 
known $\conexp^{\np}$ bound.  We also show that the membership problem for finite matrix monoids is  \pspace-complete, improving the known \nexp-upper bound.
Our two complexity 
 results are corollaries of a new polynomial bit-size bound on  matrix entries  in  finite  monoids.
This is obtained by reduction to the case of matrix groups, using the structure theory of noncommutative algebras
and of matrix monoids.
Our techniques also give us a polynomial-time algorithm for deciding whether a monoid of rational matrices is conjugate to a monoid
of integer matrices.
\end{abstract}

\newpage

\section{Introduction}
\subsection{Main Results}
The study of decision problems for matrix monoids---and their
connections with automata theory---has a long history, spanning more
than sixty years;
see~\cite{pin2010mathematical,BerstelR10} for a comprehensive survey.
The connection between automata and monoids arises from the fact that
the transition maps of automata give rise to matrix monoids with
entries in a semiring.  A particularly important instance is weighted
automata over the field $\Q$ of rational numbers.  In this setting,
questions about the support and the growth of the formal series
recognised by weighted automata reduce to questions about finitely
generated matrix monoids over~$\Q$.  In this paper we revisit several
classical decision problems for monoids of rational matrices and
obtain new complexity results.

We consider the following three problems, in which the input monoid is specified with a positive integer $d$ and a finite set of matrices $X \subseteq \mat_d(\Q)$ generating a monoid $X^*$.
\begin{enumerate}
\item \label{pb:finiteness} {\bf Finiteness Problem}: given $X\subseteq \mat_d(\mathbb Q)$, determine whether $X^*$ is finite.

\item \label{pb:membership} {\bf Finite Membership Problem}: given $X\subseteq \mat_d(\mathbb Q)$ and a target matrix $a$, determine whether $a \in X^*$, under the promise that $X^*$ is finite.
\item \label{pb:integrality} {\bf Integrality Problem}: given $X\subseteq \mat_d(\mathbb Q)$, determine whether $X^*$ is conjugate to a matrix monoid with integer entries.
\end{enumerate}

Our first main result shows that the Finiteness Problem is in \pspace, improving the $\conexp^{\np}$ upper bound of~\cite[Theorem~13]{BumpusHKST20}.
Our second main result establishes that the Finite Membership Problem is \pspace-complete.
The \pspace upper bound improves the \nexp bound of~\cite[Theorem~14]{BumpusHKST20}, and we provide a matching \pspace lower bound.
A key technical ingredient underlying both upper bounds is a polynomial bound (in the input size of the generators) on the bitsize of matrix entries appearing in finite monoids.
Using tools developed for these two results, we further show that both the decision and function versions of the Integrality Problem can be solved in polynomial time.

\subsection{Context and Related Work}

Finiteness is decidable in polynomial time for subgroups of the group $\gl_d(\mathbb Q)$ of invertible matrices in $\mat_d(\mathbb Q)$.
If a finite subgroup has generators with integer entries of absolute value at most $M$, the entries of every matrix in the group have 
absolute value at most
$M^{d^2}d^{3d^2+3}$~\cite[Theorem 1.1]{BabaiFinitenessGroups}.
Finiteness is also decidable in polynomial time for finitely generated 
submonoids of $\mat_d(\mathbb N)$.  If such a monoid is finite then it consists
of matrices whose entries are at most $M^{d-1}5{d/2}d^d$~\cite[Theorem A.2]{WeberS91}, where $M$ is the maximum entry of a generator of the monoid.

The decidability of the Finiteness Problem for matrix monoids over the rationals was first shown by McNaughton and Zalcstein~\cite{mcnaughton1975burnside}, who proved that a finitely generated matrix monoid is infinite if and only if it contains an element of infinite order.
This characterization implies that infiniteness is recursively enumerable and hence that finiteness is decidable (since finiteness is also recursively enumerable).
Subsequently, Mandel and Simon~\cite{mandel1977finite} and Jacob~\cite{jacob1977algorithme} gave effective, albeit non-elementary, bounds on the cardinality of a finitely generated submonoid of $\mat_d(\Q)$ in terms of the dimension $d$ and the number of generators.

The first elementary bound was obtained only recently, in~\cite{BumpusHKST20}, which shows that the cardinality is bounded by $m^{2^{O(d^2 \log d)}}$, where $d$ is the dimension and $m$ is the number of generators.
This cardinality bound was obtained from a \emph{diameter bound}~\cite[Theorem~1]{BumpusHKST20}:
every element of a finite monoid can be expressed as a product of generators of length at most $2^{d(2d+3)} g(d)^{d+1}$, where
$g(d) \leq 2^d d!$ is the maximum order of a finite subgroup of $\gl_d(\Q)$.
This bound is obtained using  multilinear algebra and a graph construction, previously used to compute the Zariski closure of a finitely generated matrix monoid~\cite{HrushovskiOP018}.
In this paper we reformulate the main technical result of~\cite{BumpusHKST20} in terms of factorisation forests, from which the diameter bound follows directly.

On the other hand,
the non-elementary  cardinality bounds of~\cite{jacob1977algorithme,mandel1977finite} proceed by reducing the general case to that of \emph{irreducible monoids}.
Recall that a matrix monoid $M \subseteq \mat_d(\Q)$ is irreducible if the only subspaces of $\Q^d$ invariant under all matrices in $M$ are $\set{0}$ and $\Q^d$.
Sch\"utzenberger showed that such a monoid, if finite, has cardinality at most $(2d+1)^{d^2}$~\cite{sch1962finite}.  This  bound has very recently been improved to $3^{d^2}$, with an almost matching lower bound of $2^{\lfloor  {d^2}/ 4\rfloor}$~\cite{KR2026,steinberg2026}.
To lift bounds from irreducible monoids to general monoids, Berstel and Reutenauer use a change of basis that simultaneously puts all matrices in  upper block triangular form, with square diagonal blocks, 
such that the monoid generated by the matrices appearing in each diagonal block is irreducible~\cite[Chapter~9, Theorem~1.1]{BerstelR10}.
This lifting argument depends only on the \emph{existence} of such a change of basis, not on its effective computation.
Indeed, Berstel and Reutenauer~\cite{BerstelR10} posed as an open problem the decidability of irreducibility, \ie whether the above normal form consists of a single block. 
In fact, a procedure for determining irreducibility had already been given by~\cite{ronyai1992algorithmic}.  Nevertheless, determining irreducibility and computing the above-mentioned  upper block triangular form
is still not known to be doable in polynomial time.

The improved cardinality bounds and complexity results in this paper
exploit a combination of a suitable variant of the approach
of~\cite{BerstelR10} with the diameter bound of~\cite{BumpusHKST20}.
We observe that the explicit bounds for irreducible finite matrix
monoids extend to the broader class of monoids whose enveloping
algebra is semisimple.  We then lift these bounds to general matrix
monoids by combining (i) an upper block triangular form based on the
Wedderburn--Malcev decomposition of the enveloping algebra as a direct
sum of a semisimple algebra and a nilpotent ideal with (ii) the
diameter bound of~\cite{BumpusHKST20}.  In contrast to the reduction
to upper block triangular form with irreducible blocks on the
diagonal, as used in~\cite{BerstelR10}, the polynomial-time
computability of the Wedderburn--Malcev decomposition is
straightforward.  For our approach it is essential that this
decomposition admit a polynomial-size representation.

\section{Preliminaries}

We write $\mat_{d,e}(R)$ for the set of $d \times e $ matrices with entries in a ring $R$ and denote by $\mat_d(R) := \mat_{d, d}(R)$  the set of square matrices of dimension $d$.  We denote the $d \times d$ identity matrix by $\id_d$. Given $X,Y \subseteq \mat_d(R)$, we write $XY \subseteq \mat_d(R)$ for the element-wise product and, 
for $r\in \N$, we write $X^r$ for the $r$-fold product of $X$ with itself.
Note that $X^0 = \{ \id_d\}$.
We also write \(X^{\le r}:=\bigcup_{i=0}^r X^i\) and \(X^{<r}:=\bigcup_{i=0}^{r-1}X^i\).  The submonoid of \(\mat_d(R)\) generated by $X$ is \( X^*:= \bigcup_{i=0}^\infty X^i\), while the subsemigroup
generated by $X$ is \(X^+:=\bigcup_{i=1}^\infty X^i\).

For an invertible matrix $b$, we denote by  $\cof(b)$  the cofactor matrix of $b$, that is, the $(i,j)$-entry of $\cof(b)$ is \((-1)^{i +j} \det(c(i,j))\)
where $c(i,j)$ denotes the matrix obtained from $b$ by deleting row $j$ and column $i$. Then  Cramer's rule gives \(b^{-1} = \frac{1}{\det(b)} \cof(b)^T\).

\subsection{\Tmagn of Matrices}

    Given a finite set of matrices $X\subseteq \mat_{m,n}(\Q)$, and an integer $q$, we say that $q$ is a \emph{common denominator} of $X$ if  \(q \cdot X \subseteq  \mat_{m,n}(\Z)\).
    The \emph{\tmagn} of $X$ is defined as:
    $$\magn X=\max(m,n)\cdot q\cdot p$$
    where $q$ is the least common denominator of $X$ and $p$ is the maximum absolute value of entries of elements of $qX$.

  In particular, for a reduced fraction $\frac p q$ seen as an element of $\mat_1(\Q)$ we have $\magn{\frac p q}=|pq|$.
  The main idea is that the log of the \tmagn corresponds to (an upper bound of) the number of bits needed to represent entries of a matrix.

The height is submultiplicative with respect to matrix multiplication and 
satisfies a form of triangle inequality. Let 
$X\subseteq \mat_{m,n}(\Q)$ and $Y \subseteq \mat_{n,\ell}(\Q)$ be finite sets of matrices. Then:
  \begin{gather}
         \magn{XY}\leq \magn X\magn Y \qquad \text{and} \qquad \magn{\Sigma_{i=1}^kX}\leq k\magn X.\label{obs:triangle}
  \end{gather}

Indeed, observe that given $x_1,\ldots,x_k\in X$, the least common denominator of $(x_1+\ldots+x_k)$ divides the one of $X$. The triangle inequality for $|\cdot|$ then gives the property.

    We relate below the \tmagn of a matrix with its inverse, which will prove useful:
\begin{claim}
    \label{claim:inverse-mag}
        Let $d\geq 1$ and let $a\in \gl_d(\Q)$, then $\magn{a^{-1}}\leq \magn{a}^{2d}$.
    \end{claim}

    \begin{proof}
        Let $q$ be the least common denominator of $a$ and let $p$ be the maximum absolute value of entries of $b=qa$.
        By Cramer's rule, $b^{-1}=\frac{1}{\det(b)}\cof(b)^T$.
        From the Hadamard bound, $\det(b)\leq d^{\frac d 2}p^d$.
        For the same reason, we have $\magn {\cof(b)}\leq d d^{\frac d 2}p^d$.
        Thus 
        \[\magn{b^{-1}} \, \leq\, \det(b) \, \magn {\cof(b)} \, \leq \, dd^d p^{2d}\,.\]
        Finally since $a^{-1}=qb^{-1}$, we have by the scalar submultiplicative property that 
        \[\magn{a^{-1}} \, \leq \, q \magn {b^{-1}}\, \leq \, qd^{d+1}p^{2d}\, \leq \, \magn a^{2d} \,. \qedhere\]
    \end{proof}

\subsection{Semisimplicity and the Jacobson Radical}
\label{sec:semisimple}

A vector space \(A \subseteq \mat_d(\mathbb{Q})\) containing $\id_d$ is called an \emph{algebra} if \(A\) is closed under matrix multiplication. Given \(X \subseteq \mat_d(\mathbb{Q})\), the vector space spanned by \(X^{\ast}\), or equivalently the smallest algebra \(E\) containing  \(X\) is called the \emph{enveloping algebra} \(\enve(X)\). For example,
let 
\[
a :=
\begin{pmatrix}
1 & 1 \\
0 & 0
\end{pmatrix},
\qquad
b :=
\begin{pmatrix}
0 & 1 \\
0 & 1
\end{pmatrix},
\qquad
X := \{a,b\}.
\]
Recall that $X^*$ denotes the submonoid generated by $X$, which contains the
identity matrix $\id_2$. Direct computation gives
\[
a^2 = a, \qquad b^2 = b, \qquad
ab =
\begin{pmatrix}
0 & 2 \\
0 & 0
\end{pmatrix},
\qquad
ba = 0.
\]
 Hence \(
X^*
=
\left\{
\id_2,\,
0,\,
a,\,
b,\,
ab
\right\}\).
Moreover, \(
ab=a+b-\id_2\), 
and therefore
\[
\enve(X)  = \spn_\Q\left\{a,b, \id_2\right\} 
 = \left\{
\begin{pmatrix}
x & y \\
0 & z
\end{pmatrix}
\;\middle|\;
x,y,z \in \mathbb{Q}
\right\}.
\]

A key notion when studying algebras is that of  (nilpotent) ideals. A vector subspace \(I \subseteq E\) is called an \emph{ideal} if \(E \cdot I=I=I \cdot E\).
An ideal \(I\) is called \emph{nilpotent} if there exists $d\in \N$ such that $I^d = \{0\}$; the smallest such  exponent $d$ is called the nilpotency index of $I$. In the above example, the sets
\begin{align*}
    &I_1=\left \{ 
    \begin{pmatrix}
        x & y \\ 
        0 & 0
    \end{pmatrix} \mid x,y \in \mathbb{Q} \right \}
    && I_2=\left \{ 
    \begin{pmatrix}
        0 & y \\ 
        0 & z
    \end{pmatrix} \mid y,z \in \mathbb{Q} \right \}
    && I_3=\left \{
    \begin{pmatrix}
        0 & y \\ 
        0 & 0
    \end{pmatrix} \mid y \in \mathbb{Q}\right \}
\end{align*}
are ideals, with \(I_3\) being a nilpotent ideal of nilpotency index \(2\). Nilpotent ideals are a cause of complexity, but thankfully sets of strictly upper block triangular matrices like \(I_3\) are (up to conjugation) the only examples of nilpotent ideals. In the following we repeat a proof of this fact, as it will be crucial in our size bounds for finite monoids, but we require some development first. 

The trace \(\mathrm{tr}(a)\) of \(a \in \mat_d(\mathbb{Q})\) is the sum of the diagonal entries of \(a\). An algebra \(E \subseteq \mat_d(\mathbb{Q})\) is called \emph{semisimple} if \(\{0\}\) is its only nilpotent ideal. We define \(J(E):=\{a \in E \mid \forall b \in E \colon \mathrm{tr}(ab)=0\}\) and have the following classical theorem of Dickson~\cite{dickson1924algebras}.

\begin{theorem}
    The set $J(E)$ is a nilpotent ideal and  the quotient
    $E/J(E)$ is semisimple. 
\label{thm:dickinson}
\end{theorem}

Theorem \ref{thm:dickinson} implies that \(J(E)\) is the largest nilpotent ideal, that is,  $J(E)$ is identical to  the Jacobson radical of $E$. In Section \ref{sec:filtration} and Section \ref{sec:wedderburn-malcev} we will recall a proof of the fact that \(J(E)\) can be \emph{effectively} conjugated to strictly upper block triangular matrices, hence establishing that indeed nilpotent ideals are strictly upper block triangular. To this end, we have to compute a basis of \(J(E)\) as follows. Let $B=\{b_1,\ldots,b_k\}\subseteq \mat_d(\mathbb Q)$, with $k\leq d^2$,
be a basis of $E$.
Define the $\mathbb Q$-linear \emph{trace map} \(\tmap_B \colon E \to \mathbb Q^k\) by
\[\tmap_B(a) := (\tr(ab_1), \ldots, \tr(ab_k))\, . \]
With respect to the basis $B$ on $E$ this map is represented by
the \emph{trace matrix}

\begin{equation}\label{eq: trace_mat}
\tmatrix_B=(\tr(b_ib_j))_{1\leq i,j \leq k } 
\end{equation}
We have \(ker(\tmap_B)=J(E)\): Indeed, since the trace \(\mathrm{tr}\) is a linear map, \(\mathrm{tr}(ab_1)= \dots= \mathrm{tr}(ab_k)=0\) implies for all \(b=\sum_{i=1}^k \lambda_i b_i \in E\) that
\[\mathrm{tr}(ab)=\mathrm{tr}\left(a\sum_{i=1}^k \lambda_i b_i \right )=\sum_{i=1}^k \lambda_i \mathrm{tr}(ab_i)=0.\]

\subsection{Block Decompositions of Matrices}
\label{sec:filtration}
A \emph{filtration} of the vector space $\mathbb Q^d$ is a strictly increasing sequence 
of subspaces
\begin{gather}
    \{0\}=W_0 \subset W_1 \subset \cdots \subset W_k = \mathbb Q^d \, .
\label{eq:FILT}
\end{gather}
We say that a basis $B=\{v_1,\ldots,v_d\}$ of $\mathbb Q^d$ is \emph{adapted to the filtration}~\eqref{eq:FILT}
if it arises as the disjoint union of bases of 
subspaces $V_1,\ldots,V_k \subseteq \mathbb Q^d$
such that $W_i = V_1 \oplus \cdots \oplus V_i$ for $i \in \{0,\ldots,k\}$. Such an adapted basis can be constructed by first choosing a
basis of \(W_1\), and then successively extending a basis of \(W_i\) to a
basis of \(W_{i+1}\), for \(i=1,\ldots,k-1\).
An algebra $E\subseteq \mat_d(\mathbb Q)$ is said to \emph{preserve the filtration}~\eqref{eq:FILT} if $E \cdot W_i \subseteq W_i$
for all $i \in \{0,\ldots,k\}$.  
In such a case, let $p\in \mat_d(\mathbb Q)$ be the matrix whose columns are given by $B$.
Then for all $b \in E$ the matrix $a:=p^{-1}bp$ has the upper block triangular form
\[
    a=\begin{pmatrix}
a_{1,1} & a_{1,2} & \dots & a_{1,k} \\
0 & a_{2,2} & \dots & a_{2,k} \\
\vdots & 0 & \ddots & \vdots \\
0 & \dots & 0 & a_{k,k}
\end{pmatrix},
\] 
where \(a_{i,j}\) has dimension $d_i \times d_j$ for \(j\geq i\), where $d_i =\dim(V_{i})$ and 
$d_j=\dim(V_{j})$.
Here we  say that $(d_1,\ldots,d_k)$ is the \emph{signature} of the block decomposition of  $a$. 

Note that if $E=\mathrm{Env}(X)$ is the enveloping algebra of
a set of matrices $X\subseteq \mat_d(\mathbb Q)$, then for an invertible matrix $p\in \mat_d(\mathbb Q)$, the conjugated algebra $p^{-1}Ep$
is in upper block triangular form if and only if $p^{-1}Xp$ is upper block triangular.

\section{Structure Theory of Matrix Monoids and Algebras}
In this section we give convenient formulations of known results concerning, respectively, the structure of matrix monoids and their enveloping algebras, that are key ingredients of the main results of this paper.  

\subsection{Wedderburn--Malcev Form}
\label{sec:wedderburn-malcev}
We say that an algebra $E \subseteq \mat_d(\mathbb Q)$ is in \emph{Wedderburn--Malcev form} if, with respect to  some signature,
$E$ consists of upper block triangular matrices and 
the Jacobson radical $J(E)$ consists of  strictly upper block triangular matrices.  In this case we have that
\( \BlockDiag(E) \cong E/J(E)\)
is semisimple, where $\BlockDiag(\cdot)$ is the projection onto the set of block diagonal matrices.%
\begin{proposition}
\label{prop:trig}
    Let $E$ be a sub-algebra of $\mat_d(\mathbb Q)$ and let $J$ be a nilpotent ideal in $E$.
    Given a  basis $B$ of \(J\), one can compute in polynomial time an
    invertible matrix $p\in \mat_d(\mathbb Q)$ such that 
    \begin{description}
        \item[(i)] the conjugated algebra $p^{-1}Ep$ and conjugated ideal  $p^{-1}Jp$ consist of upper block triangular and strictly upper block triangular matrices, respectively;
        \item[(ii)]   $\magn{p}\leq \magn B^{d}$ and $\magn{p^{-1}}\leq \magn B^{2d^2}$.
    \end{description}
\end{proposition}
\begin{proof}
We first show Item~(i).
    Let $k\leq d$ be the nilpotency index of $J$ and write $W_i := J^{k-i}\mathbb Q^d$.  Then we obtain the following filtration:
    \begin{gather}
     \{0\}=W_0 \subset W_{1} \subset \cdots \subset W_k = \mathbb Q^d\, .
\label{eq:filtration}
\end{gather}
    The above chain of inclusions is strict. Indeed, if the sequence were not strict then we would have $W_0=W_i$ for some $i<k$.
    But this contradicts the fact that $W_0=\{0\}$ and $W_i \neq \{0\}$ by minimality of $k$.
Note, moreover, that $E$ preserves the filtration~\eqref{eq:filtration}.  
Indeed, since $J$ is an ideal, we have \(E\cdot W_i = E \cdot J^{k-i} \mathbb Q^d  \subseteq J^{k-i} \mathbb Q^d = W_i \, \).

    Let $\{v_1,\ldots,v_d\}$ be a basis adapted to the filtration in~\eqref{eq:filtration}, as defined in Section~\ref{sec:filtration}, and 
    let $p$ be the matrix whose $i$-th column is $v_i$, for $i\in \set{1,\ldots,d}$. 
Since $E$ preserves the filtration, it follows that  the conjugated algebra $p^{-1}Ep$ consists of  upper block triangular matrices. 
We claim furthermore that 
the conjugated ideal  $p^{-1}Jp$ consists of  strictly upper block triangular matrices.  To see this,
let  $a\in J$  and $i\ge1$. Then the inclusion
   $ a\, W_i \subseteq J\, W_i= W_{i-1}$ holds, meaning that multiplication by $a$ strictly
lowers the index in the filtration.
This implies that $p^{-1} a p$ is strictly upper block triangular in
the adapted basis.

    \Tmagn analysis: We next establish Item~(ii). We start by computing bases $B_i$ of the $W_is$. We take $B_k$ to be the canonical basis of $W_k$.
    For $i\in \set{1,\dots, k}$, $B_iB$ spans $W_{i-1}$. By submultiplicativity of the \tmagn we obtain bases such that  $\magn{B_i}\leq \magn{B}^d$ for any $i\in \set{1,\ldots,k}$.
    From these bases we can select a basis adapted to the filtration in~\eqref{eq:filtration}: first select a basis of $W_1$. Then for $i\in \set{1,\ldots, k-1}$, given a basis adapted to the filtration $W_0\subset\ldots \subset W_{i}$, we can extend it to a basis adapted to $W_0\subset \ldots \subset W_{i+1}$ by adding vectors from $B_{i+1}$.
    
    Let $p$ be the change of basis matrix obtained from these vectors,
    then $\magn p \leq \magn B ^d$. Using Claim~\ref{claim:inverse-mag}, above, we obtain $\magn {p ^{-1}}\leq \magn B^{2d^2}$.
\end{proof}

\begin{theorem}
There is a polynomial-time algorithm that, given a 
basis \(B\) of an algebra $E \subseteq \mat_d(\Q)$,
determines an invertible matrix \(p \in \mat_d(\Q)\) such that \(p^{-1}Ep\) is in Wedderburn--Malcev form.
Moreover we have  $\magn {\set{p,p^{-1}}}=\magn B ^{O(d^4)}$ .

\label{TheoremTraceNormalFormComputation}
\end{theorem}
\begin{proof}
By bilinearity of the trace, the Jacobson radical $J(E)$ is defined by the following system of linear equations:
\[ J(E) = \{a \in E \mid \forall b \in B \colon \mathrm{tr}(ab)=0\} \, .\]
Thus we can compute a basis $B'$ of $J(E)$ from $B$ in polynomial time.

Applying Proposition~\ref{prop:trig} to $E$ and $J:=J(E)$, given in terms of the basis $B'$, 
we can compute in polynomial time an invertible matrix $p \in \mat_d(\mathbb Q)$ such that
$p^{-1} E p$ is  upper block triangular and~$p^{-1} J p$ is strictly upper block triangular.

    \Tmagn analysis.
    Using the Bareiss algorithm, the Hadamard bound gives us a bound on the \tmagn of elements of $B'$. This is similar to the result of Claim~\ref{claim:inverse-mag}, except that here the dimension is bounded by $d^2$: $\magn {B'}\leq \magn B ^{O(d^2)}$. 

    Applying  Proposition \ref{prop:trig}, we obtain $\magn {\set{p,p ^{-1}}}=\magn B^{O(d^4)}$.
\end{proof}

\subsection{Diameter Bound for Finite Monoids via Factorisation Forests}
\label{sec:forest}
Recall that a finitely generated monoid has diameter $D$ if every element of the monoid can be written as the product of a list of 
generators of length at most $D$.
The diameter bound for finitely generated matrix monoids in~\cite{BumpusHKST20} is proved using a graph construction
that was introduced in~\cite{HrushovskiOP018} for computing the Zariski closure of matrix monoids.  Below, we reformulate their
main technical contribution in terms of factorisation forests, extracting a result that we think is of independent interest.

Given a morphism $\varphi:\Sigma^+\rightarrow \mat_d(\mathbb Q)$, a \emph{Ramseyan factorisation forest}
of a word $w\in \Sigma^+$ is a finite unranked plane tree whose nodes are labelled with elements of $\Sigma^+$.
The root is labelled $w$, the children have labels in $\Sigma$, and an internal node with label $x$
whose children are labelled $x_0,\ldots,x_{s-1}$, satisfies 
\begin{description}
    \item[(i)] $\varphi(x)=\varphi(x_0x_1\cdots x_{s-1})$, and
    \item[(ii)] if $s\geq 3$ then 
$\varphi(x_0),\ldots,\varphi(x_{s-1})$ generate a \emph{subgroup} of $\mat_d(\mathbb Q)$. 
\end{description}
The \emph{height} of such a tree is the maximum number of non-leaves on a root-to-leaf path and the
\emph{reduced height} is the maximum number of nodes of arity at least 3 on a root-to-leaf path.

The above notion of factorisation forest differs from the classical notion, due to Simon~\cite{Simon90}, which involves the stronger condition 
that for every node with 3 or more children, that node and all its children should be  the same idempotent element of the monoid.
The payoff is that we obtain a factorisation forest whose height depends only on the dimension of the matrices, not the cardinality of the monoid (as is the case for the classical formulation).

\begin{theorem}\label{thm:factorization}
  If $\varphi$ has finite image then every word $w\in \Sigma^+$ has a Ramseyan factorisation forest of height at most
$2d(d+3)$ and reduced height at most $d$.
\end{theorem}

We first show  that  Theorem \ref{thm:factorization} implies the claimed diameter bound, before proceeding to its proof.

\begin{corollary}\cite[Theorem 1]{BumpusHKST20}
Let $d$ be a positive integer and let $X\subseteq \mat_d(\mathbb Q)$.
If $X^*$ is finite, then its diameter is at most $2^{2d(d +3)}g(d)^d$, where $g(d)$ is the maximum cardinality
of a subgroup of $\mat_d(\mathbb Q)$.
\label{cor:length}
\end{corollary}
\begin{proof}
Let $\varphi:\Sigma^+\rightarrow \mat_d(\mathbb Q)$ be a morphism such that $\varphi(\Sigma)=X$. With respect to $\varphi$,
every word $w\in \Sigma^+$ has a Ramseyan factorisation forest of height at most $2d(d+3)$ and reduced height at most $d$.
We can moreover assume that the non-binary nodes of such a tree have arity at most $g(d)$.  Such a tree thus has at most
$2^{2d(d +3)}g(d)^d$ leaves.  The result follows.
\end{proof} 

The starting point for proving Theorem \ref{thm:factorization} is a
factorisation forest theorem for general matrix monoids (both finite
and infinite) in~\cite[Proposition 2]{el2026algebraic}. We then
transform such a factorisation forest into
a Ramseyan factorisation forest,  using a construction similar to~\cite[Lemma 11]{BumpusHKST20}.  
Given a morphism $\varphi:\Sigma^+\rightarrow \mat_d(\mathbb Q)$, a \emph{weak  factorisation forest}
of a word $w\in \Sigma^+$ is a finite unranked plane tree whose nodes are labelled with elements of $\Sigma^+$.
The root is labelled $w$, the children have labels in $\Sigma$, and a node with label $x$
whose children are labelled $x_0,\ldots,x_{s-1}$, satisfies
\begin{description}
    \item[(i)] $\varphi(x)=\varphi(x_0x_1\cdots x_{s-1})$, and
    \item[(ii)] if $|s|\geq 3$ then 
$\rk(\varphi(x_0))= \cdots= \rk(\varphi(x_{s-1})) = \rk(\varphi(x)) = \rk(\varphi(x)^2)$. 
\end{description}

The height and reduced height of a weak factorisation forest are defined exactly as in the Ramseyan case.
The following result gives bounds on the height and reduced height of weak factorisation forests in arbitrary matrix monoids.
\begin{theorem}[\cite{el2026algebraic}, Proposition~3]\label{thm:weakfactorisation}
    Let $\varphi : \Sigma^+ \to \mat_d(\Q)$ be a semigroup morphism. Then every word has a weak  factorisation forest of height $\leq d(d+3)$ and reduced height $\leq d$. 
\end{theorem}

Recall that the notion of Ramseyan factorisation forest strengthens Condition~(ii) to the requirement that $\varphi(x_0), \ldots, \varphi(x_{s-1})$ generate a subgroup of $\mat_d(\Q).$
We next show how to transform a weak factorisation forest to a Ramseyan factorisation forest in case the morphism $\varphi$ has finite image.
To this end, we use the following result, a direct consequence of~\cite[Proposition 9]{HrushovskiOP018} and~\cite[Lemma 7]{BumpusHKST20}.
\begin{lemma}\label{lemma:boundprod}
    Let $a_1, \ldots, a_\ell \in \mat_d(\Q)$ and $a := a_1 \cdots a_\ell$. Assume that
    \begin{equation}
    \label{eq-rank}
        \rk(a_1) = \cdots = \rk(a_\ell) = \rk(a)
    \end{equation}
    Then there exist $1 = i_1 < i_2 < \cdots < i_t = \ell$, where $t\leq 2^d+1$, such that
    $\ker(a)=\ker(a_{i_1}\cdots a_{i_t})$ and  
    $\Imag(a)=\Imag(a_{i_1}\cdots a_{i_t})$.
\end{lemma}

    The following is a version of~\cite[Lemma 11]{BumpusHKST20}.
\begin{proposition}\label{pro:reform}
   Assume $\varphi$ has a finite image. Let $s\geq 3$ and
   $w,w_0,\ldots,w_{s-1} \in \Sigma^+$ such that:
   \begin{itemize}
       \item~$\varphi(w) = \varphi(w_0w_1\cdots w_{s-1})$, 
       \item~ $ \rk(\varphi(w_0)) = \cdots = \rk(\varphi(w_{s-1})) =  \rk(\varphi(w)) = \rk(\varphi(w)^2)$.
    \end{itemize}
    Then there exist $u_0, \ldots, u_{t -1} \in \{ w_0 , \ldots, w_{s -1}\}^{\leq 2^{d +2}}$ such that $\varphi(w) = \varphi(u_0 \cdots u_{t -1})$ and, moreover, $\varphi(u_0), \ldots, \varphi(u_{t -1})$ generate a subgroup of $\mat_d(\Q).$ 
\end{proposition}
\begin{proof}
For $i \in \{0,\ldots,s-2\}$,
we apply Lemma~\ref{lemma:boundprod} to $\varphi(w_{i + 1}) \varphi(w_{i +2}) \cdots \varphi(w_{s -1})$ and $\varphi(w_0)\varphi(w_1) \cdots \varphi(w_i)$ respectively to obtain $x_i, y_i \in \{ w_0 , \ldots, w_{s -1}\}^{\leq 2^d + 1}$ such that 
\begin{itemize}
\item  $\Imag(\varphi(x_i)) = \Imag(\varphi(w_{i + 1})), \quad \ker( \varphi(x_i)) = \ker(\varphi(w_{s -1}))$.
\item $\Imag(\varphi(y_i)) = \Imag(\varphi(w_0)), \quad  \ker( \varphi(y_i)) = \ker(\varphi(w_i)).$
\end{itemize}

From the rank condition $ \rk(\varphi(w_0)) = \cdots = \rk(\varphi(w_{s-1})) =  \rk(\varphi(w)) = \rk(\varphi(w)^2),$ we obtain that $\ker (\varphi(y_i)) \cap \Imag(\varphi(x_i)) =\{0\}  $, which implies that 
\[
\ker(\varphi(y_ix_i )) = \ker (\varphi(x_i)) = \ker(\varphi(w_{s -1})), \quad \Imag( \varphi(y_ix_i) )= \Imag(\varphi(y_i)) = \Imag(\varphi(w_0)).
\]
Let $p$ be the smallest integer such that $\varphi(y_ix_i)^{p} = \varphi(y_ix_i)^{2p}$ for all $i$. Such an integer $p$  exists since $\varphi$ has a finite image. Moreover, $\varphi(y_ix_i)^{p}$ acts as the identity on $\Imag(\varphi(y_ix_i)) = \Imag(\varphi(w_0))$.
We then rewrite  $\varphi(w)$ as
    \[
\varphi(w)
= \varphi(w_0 x_0)\,
   \varphi(y_0 x_0)^{p-1}\,
   \varphi(y_0 w_1 x_1)\,
   \varphi(y_1 x_1)^{p-1} \cdots
   \varphi(y_{s-2} x_{s-2})^{p-1} \,  \varphi(y_{s-2} w_{s-1}).
\]
 Let $(u_j)_{j=0}^{t -1}$ be the finite sequence obtained by listing the factors \((u_j)_{j=0}^{t -1} :=\)
\[\Big(
w_0x_0,\;
\underbrace{y_0x_0,\ldots,y_0x_0}_{p-1\text{ times}},\;
y_0w_1x_1,\;
\underbrace{y_1x_1,\ldots,y_1x_1}_{p-1\text{ times}},\;
\ldots,\;
\underbrace{y_{s-2}x_{s-2},\ldots,y_{s-2}x_{s-2}}_{p-1\text{ times}},\;
y_{s-2}w_{s-1}
\Big).
\]
Then  each $u_j$ belongs to $\{ w_0 , \ldots, w_{s -1}\}^{\leq 2^{d +2}},$ and $\varphi(w)=\varphi(u_0\cdots u_{t-1})$.

Let $G = \{ \varphi(u_0), \ldots,  \varphi(u_{t -1})\}^* \subseteq M_d(\mathbb Q)$. Note that for every $0 \leq i \leq t-1$,  $u_i$ has one of the forms $w_0 x_0, y_j x_j, y_jw_{j+1}x_{j +1}, y_{s -2}w_{s-1}$. 
Therefore, for every $0 \leq i \leq t-1$, $\ker(\varphi(u_i)) = \ker( \varphi(w_{s -1}))$, $\Imag(\varphi(u_i)) = \Imag(\varphi(w_0))$, and moreover, $\ker(\varphi(u_i)) \cap \Imag(\varphi(u_i)) = \{0\}$. 
Since the $\varphi(u_i)$  all have the same kernel $U$ and image $V$ such that moreover $U\cap V=\{0\}$, $G$ is isomorphic to a finite monoid 
of invertible linear endomorphisms of $V$, i.e.\ is a group.
\end{proof}

\begin{proof}[Proof of Theorem~\ref{thm:factorization}]
  We will show that every $v \in \Sigma^+$ has a Ramseyan
  factorisation forest of height at most $2d(d + 3)$, and reduced
  height at most $d$.  Fix $v \in \Sigma^+$. The construction of the
  Ramseyan factorisation forest of $v$ relies on the weak
  factorisation forest given in
  Theorem~\ref{thm:weakfactorisation}. Let $T$ be the weak
  factorisation forest of $v$, with height at most $d(d +3)$ and
  reduced height at most $d$.  We modify $T$ as follows. Process the
  nodes with more than two children level by level, starting from the
  leaves and moving upward. Whenever a node labelled by $w$ has more
  than two children labelled $w_0, \ldots, w_{s -1}$, apply
  Proposition~\ref{pro:reform} to that node, and replace that node by
  the corresponding subtree as in Figure~\ref{fig:tree}.
\begin{figure}[!h]
\centering
\scalebox{0.8}{
\begin{tikzpicture}[
  every node/.style={font=\small},
  brace/.style={decorate,decoration={brace,amplitude=4pt},red,thick}
]
\usetikzlibrary{decorations.pathreplacing}

\node (w)  at (8.5,4.2) {$w$};
\node (u0) at (2.0,3.0) {$u_0 = w_0x_0$};
\node (u1) at (6.5,3.0) {$u_1 = y_0x_0$};
\node (ud) at (8.5,3.0) {$\cdots$};
\node (ut) at (14.5,3.0) {$u_{t-1} = y_{s-1}w_{s-1}$};

\draw (w) -- (u0);
\draw (w) -- (u1);
\draw (w) -- (ud);
\draw (w) -- (ut);

\node (w0)  at (1.0,2.0) {$w_0$};
\node (x0a) at (3.0,2.0) {$x_0$};
\draw (u0) -- (w0);
\draw (u0) -- (x0a);

\node (y0)  at (6.5,2.0) {$y_0$};
\node (x0b) at (10.0,2.0) {$x_0$};
\draw (u1) -- (y0);
\draw (u1) -- (x0b);

\node (ys1) at (13.5,2.0) {$y_{s-1}$};
\node (ws1) at (15.5,2.0) {$w_{s-1}$};
\draw (ut) -- (ys1);
\draw (ut) -- (ws1);


\node (xa1) at (2.5,1.25) {$\cdot$};
\node (xa2) at (3.5,1.25) {$\cdot$};
\draw (x0a) -- (xa1);
\draw (x0a) -- (xa2);

\node (xv1) at (2.5,0.35) {$\vdots$};
\node (xv2) at (3.5,0.35) {$\vdots$};
\draw (xa1) -- (xv1);
\draw (xa2) -- (xv2);

\node (xL1) at (1.8,-0.55) {$w_{i_1}$};
\node (xLm) at (4.2,-0.55) {$w_{i_{\ell_1}}$};
\draw (xv1) -- (xL1);
\draw (xv2) -- (xLm);
\node at (3.0,-0.55) {$\cdots$};

\node (ya1) at (6.0,1.25) {$\cdot$};
\node (ya2) at (7.0,1.25) {$\cdot$};
\draw (y0) -- (ya1);
\draw (y0) -- (ya2);

\node (yv1) at (6.0,0.35) {$\vdots$};
\node (yv2) at (7.0,0.35) {$\vdots$};
\draw (ya1) -- (yv1);
\draw (ya2) -- (yv2);

\node (yL1) at (5.3,-0.55) {$w_{j_1}$};
\node (yLm) at (7.7,-0.55) {$w_{j_{\ell_2}}$};
\draw (yv1) -- (yL1);
\draw (yv2) -- (yLm);
\node at (6.5,-0.55) {$\cdots$};

\node (xb1) at (9.5,1.25) {$\cdot$};
\node (xb2) at (10.5,1.25) {$\cdot$};
\draw (x0b) -- (xb1);
\draw (x0b) -- (xb2);

\node (xw1) at (9.5,0.35) {$\vdots$};
\node (xw2) at (10.5,0.35) {$\vdots$};
\draw (xb1) -- (xw1);
\draw (xb2) -- (xw2);

\node (kL1) at (8.8,-0.55) {$w_{i_1}$};
\node (kLm) at (11.2,-0.55) {$w_{i_{\ell_1}}$};
\draw (xw1) -- (kL1);
\draw (xw2) -- (kLm);
\node at (10.0,-0.55) {$\cdots$};

\node (sa1) at (13.0,1.25) {$\cdot$};
\node (sa2) at (14.0,1.25) {$\cdot$};
\draw (ys1) -- (sa1);
\draw (ys1) -- (sa2);

\node (sv1) at (13.0,0.35) {$\vdots$};
\node (sv2) at (14.0,0.35) {$\vdots$};
\draw (sa1) -- (sv1);
\draw (sa2) -- (sv2);

\node (sL1) at (12.4,-0.55) {$w_{k_1}$};
\node (sLm) at (14.6,-0.55) {$w_{k_{\ell_3}}$};
\draw (sv1) -- (sL1);
\draw (sv2) -- (sLm);
\node at (13.5,-0.55) {$\cdots$};

\draw[brace] (0.35,-0.80) -- (0.35,3.05)
  node[midway,xshift=-6pt,left,font=\small] {$ d+2 \ge$};

\end{tikzpicture}}
\caption{A schematic Ramseyan factorisation tree for $w$ with leaves among $w_i$ of height $\le d +3$. The root $w$ is decomposed into $u_0,u_1,\ldots,u_{t-1}$.
We display $u_0=w_0x_0 ,u_1=y_0x_0, \ldots, u_{t-1}=y_{s-1}w_{s-1}$. Each factor $x_i$, $y_i$, is further refined by a binary subtree  down to leaves among the $w_j$.}\label{fig:tree}
\end{figure}
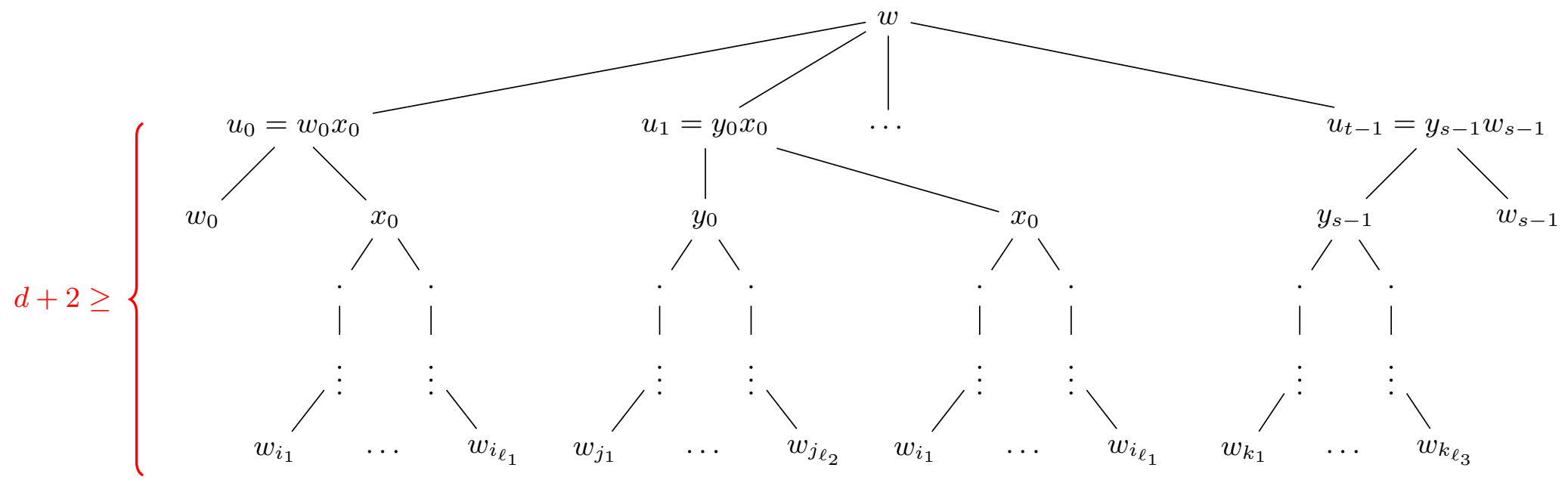
Each application of Proposition~\ref{pro:reform} increases the height below that node by at most $d+3$, and does not increase the reduced height. 
Since, for each path from the root to leaf we can pass through at most $d$ nodes that have more than two children, on every such path we can apply Proposition~\ref{pro:reform} at most $d$ times. Therefore, the total height increases by at most $d(d +3),$ and the reduced height remains unchanged.
Hence, the resulting tree has height at most $2d(d+3)$, reduced height at most $d$, and satisfies the Ramseyan factorisation forest conditions.
\end{proof}

\section{\Tmagn Bound for Finite Monoids}

In this section we prove a bound on the bitsize of the matrices in a finite matrix monoid
that is polynomial in the bitsize of the generators and the dimension $d$.  We first obtain a bitsize bound in the case that the enveloping algebra of the monoid is semisimple.  This bound combines the fact that an element of such a monoid is determined by its image under the trace map (see Section~\ref{sec:semisimple}) together with the fact that the image of the trace map is confined to a small range, due to the monoid being finite.
We then lift the bitsize bound from the semisimple case to the general case using the Wedderburn--Malcev-form computation in  Theorem~\ref{TheoremTraceNormalFormComputation} and the diameter bound in Corollary~\ref{cor:length}. We remark that similar  arguments are used  in \cite{BabaiFinitenessGroups} in the group case.

\begin{lemma}
    Let \(X^* \subseteq \mat_d(\Q)\) be a \emph{finite} monoid. Then \(\tr(a) \in \{-d,\dots,d\}\) for all \(a \in X^*\).%
    \label{LemmaSmallTrace}
\end{lemma}

\begin{proof}
    Consider a matrix \(a \in X^*\). Since \(X^*\) is finite, there exists \( m > n\) such that \(a^m = a^n\).
    It follows that the minimal polynomial of \(a\) divides \( x^n - x^m\), i.e. every eigenvalue of \(a\) is either \(0\) or a root of unity.
    In particular, \(|\lambda| \leq 1\) for all eigenvalues \(\lambda\) of \(a\) and hence \(|\tr(a)|=|\sum_{i=1}^d \lambda_i| \leq d\). 
    As a sum of algebraic integers, \(\tr(a)\) is itself an algebraic integer. Since \(\tr(a)\) is moreover rational and \(\Z\) integrally closed in $\Q$,
    we obtain \(\tr(a) \in \Z\). The lemma immediately follows.
\end{proof}


\begin{theorem}
\label{thm:finite-semisimple}
    Let $X^*\subseteq \mat_d(\Q)$ be a finite monoid such that $\enve(X)$ is semisimple, with a basis $B\subseteq X^*$. Then $\magn{X^*} = d^{O(d^2)}\magn B$.

\end{theorem}

\begin{proof}
    Let $X\subseteq \mat_d(\Q)$ be   such that $X^*$ is finite and $\enve(X)$ semisimple.  One can choose $B$ a basis of $\enve(X)$ within $X^{\leq d ^2-1}$ since $\enve(X)$ has dimension at most $d^2$ (and $\id_d\in X^*)$.
    From Equation~\eqref{obs:triangle}, $\magn{B}\leq \magn{X}^{d^2}$.

    By Lemma~\ref{LemmaSmallTrace}, the matrix $\tmatrix_B$, defined in Equation~\eqref{eq: trace_mat}, is an integer matrix with entries in $\set{-d,\ldots, d}$.
     We have by Claim~\ref{claim:inverse-mag} that $\magn {T_B^{-1}}\leq (d^3)^{2d^2}$.

    Let $a\in X^*$ and let $\alpha=(\alpha_b)_{b\in B}\in \Q^B$ be such that $a=\sum_{b\in B}\alpha_bb$. Then for any $c\in B$, $\tr(ca)=\sum_{b\in B}\alpha_b\tr(cb)=( T_B\alpha)_c$.
    By Lemma~\ref{LemmaSmallTrace}, the vector $ T_B \alpha$ has entries in $\set{-d,\ldots,d}$, and thus has \tmagn at most $d^3$.
    We write $a=\sum_{b\in B}( T_B^{-1}T_B\alpha)_bb$.
    Let $A=T_B^{-1}\set{-d,\ldots,d}^B$ and let $C$ be the set of entries of elements of $A$. Then $\magn C\leq \magn A\leq (d^3)^{2d^2+1}$.
    Then since $X^*\subseteq \sum_{b\in B}CB$, we have by triangle inequality that:
    \begin{align*}
       \magn{X^*}& \leq d^2\magn C \magn B\, 
       \leq \, d^2(d^3)^{2d^2+1}\magn B \, =\,  d^{O(d^2)}\magn B \qedhere
    \end{align*}
\end{proof}

\begin{proposition}
\label{prop:heightbound-wmnf}
    Let $X^* \subseteq \mat_d(\Q)$ be a finite monoid such that $\enve(X)$ is in Wedderburn--Malcev form, with a basis $B\subseteq X^*$.
Then $\magn{X^*} = d^{O(d^3)}\magn X^d\magn B^d$.

\end{proposition}
\begin{proof}
    By assumption we can write the set of generators of $\enve(X)$ as 
    $ X = \{a_1+c_1,\ldots,a_m+c_m\}\subseteq X^*$, where each $a_i$ is block diagonal and each $c_i$ 
    strictly upper triangular on blocks.  
    Write  $A:=\{a_1,\ldots,a_m\}$, $C:=\{c_1,\ldots,c_m\}$ and $D:=\{a_1,\ldots,a_m,c_1,\ldots c_m\}$.

We first obtain a bound on the \tmagn  of $D^*$.
By definition of Wedderburn--Malcev normal form, $\enve(A)=\BlockDiag(\enve(X))$ is semisimple.
Hence, by Theorem~\ref{thm:finite-semisimple}, $\magn{A^*}=d^{O(d^2)}\magn B$. Moreover, since any matrix in $(A^*C)^d$ is zero, we get that a non-zero matrix in $D^*$ contains at most $d-1$ occurrences of elements of $C$.
Hence $\magn{D^*}\leq(\magn{A^*}\magn C)^{d}=d^{O(d^3)}\magn X^d\magn B^d$.

Let $a \in  X^*$.  By Corollary~\ref{cor:length}, $a$ can be written as a product of at most 
\[ L(d):=  2^{2d( d+3) + d^2} (d!)^d=d^{O(d^2)}  \] matrices 
from the generating set $X$.  By the distributive law
we can write $a$ as the sum of
elements of $D^*$. Since any matrix in $(A^*C)^d$ is zero, most of these summands are zero, and the number of non-zero summands is bounded by $\sum_{i<d}\binom{L(d)}{i}\leq dL(d)^d=d^{O(d^3)}$.
By Equation~\eqref{obs:triangle}, summing $k$ elements of a set can at most multiply the \tmagn by $k$.
Thus \[\magn{X^*} \, \leq \, dL(d)^d\magn{D^*}\, =\,  d^{O(d^3)}\magn X^d\, \magn B^d\,.\qedhere\]
\end{proof}

    Relationship between \tmagn and bitsize.
    Before stating our main result on the height of a finite matrix monoid, we define $\bitsize (X)$ the \emph{bitsize} of a finite set $X$ of matrices: $\bitsize(X)$ is defined as the maximal number of bits necessary to write (the absolute values of) numerators and denominators of entries of elements of $X$. Thus any element of $X$ can be represented with $mn(\bitsize(X)+1)$ many bits.
  Observe that the least common denominator of $X$ is smaller than $2^{mn\bitsize(X)}$.
  In particular, if $m=n=d$, then the following relationships hold:
 
  \begin{align}
    &\bitsize(X)\leq \log(\magn X)\leq (2d^2+1) \bitsize(X) + \log d \label{obs:sizevheight}\\
    &\bitsize(X)\leq \log(\magn X)\leq \bitsize(X)+\log d, &\text{if } X \text{ has integer entries.}\label{obs:int-sizevheight}
    \end{align}

\begin{theorem} \label{thm:poly-bitsize}
    Let \(X \subseteq \mat_d(\Q)\) such that $X^*$ is finite, then
$\magn{X^{\ast}}=\magn X^{O(d^7)}$, where the constant in the big-oh notation is absolute\footnote{By an absolute constant we mean a fixed number, 
independent of any parameters}.
 As a consequence    $\bitsize(X^*)=O( d^{9}\bitsize(X))$. 
Note that if $X\subseteq \mat_d(\Z)$, then $\bitsize(X^*)=O( d^{7}(\log d +\bitsize(X)))$.
\end{theorem}
\begin{proof}
    The first step is to compute a basis $B$ of $\enve(X)$.
    Since $\enve(X)$ has dimension at most $d^2$, one can find a basis within $X^{d^2-1}$, which gives $\magn B\leq \magn X^{d^2}$.
    We then put the matrices in Wedderburn--Malcev normal form. Applying Theorem~\ref{TheoremTraceNormalFormComputation}, we obtain a change of basis matrix $p$ such that $\magn{p^{-1}(X\cup B)p}=\magn X^{O(d^6)}$.
    Finally, by applying Proposition~\ref{prop:heightbound-wmnf}, we obtain that $$\magn {X^*}\leq \magn {p}\magn {p^{-1}X^*p}\magn {p^{-1}}=\magn X^{O(d^7)}\, .$$
Combining this with Equation~\eqref{obs:sizevheight}, we obtain the polynomial size bound.
If $X$ contains only integer matrices then using Equation~\eqref{obs:int-sizevheight}, we obtain the result.
\end{proof}

\section{Finiteness and  Membership}

In this section, we derive complexity results for the Finiteness and Finite Membership Problems as corollaries of the polynomial bit-size bound on matrix entries in finite monoids, established in Theorem \ref{thm:poly-bitsize}.

\begin{proposition}\label{thm:finiteness}
    The Finiteness Problem  is decidable in \pspace.
\end{proposition}

\begin{proof}
\pspace membership is a consequence of the following nondeterministic polynomial-space algorithm for deciding infiniteness. The algorithm maintains  a matrix \(a \in X^*\), that is initially the identity. At each step, 
the algorithm nondeterministically chooses
     \(b \in X\) and makes the assignment~\(a := ab\).
If the algorithm reaches a matrix whose bitsize exceeds the bound given by Theorem~\ref{thm:poly-bitsize}, it accepts. (We note that  the constant in the
big-oh notation in Theorem~\ref{thm:poly-bitsize} is absolute.)
\end{proof}

\begin{proposition}
\label{prop-hardness}
  The Finite Membership Problem  is \pspace-complete.
\end{proposition}

\begin{proof}
   \pspace membership is a consequence of the following nondeterministic polynomial-space algorithm. The algorithm maintains  a matrix \(a \in X^*\), 
   that is initially the identity. At each step, it nondeterministically guesses a generator
     \(b \in X\) and performs the update \(a:=ab\). By Theorem~\ref{thm:poly-bitsize}, the bitsize of 
$a$
 remains polynomially bounded throughout the computation. The algorithm accepts if it reaches the target matrix 
$t$.

Hardness is shown by a reduction from the intersection emptiness problem for DFA. More precisely, we prove that the Finite Membership problem  over 
$\mathbb N$ is \pspace-hard in  Section \ref{app-hardness}. We also extend \pspace-hardness to matrices over any semiring $R\neq \{0\}$ and to irreducible monoids.
\end{proof}

\section{PSPACE-hardness in Proposition~\ref{prop-hardness}}
\label{app-hardness}

Recall that the DFA intersection emptiness problem asks, 
given \(n \in \mathbb N\) and DFAs \(D_1, \dots, D_n\) over a common alphabet $\Sigma$,  whether the intersection of their languages is nonempty.

For simplicity, 
we think of each DFA $D_i$ as a weighted $\mathbb N$-automaton $(\alpha_i,(M_{\sigma,i})_{\sigma\in \Sigma},\gamma_i)$ with $d_i$ states,
where $\alpha_i \in \{0,1\}^{1 \times d_i}$ is the initial row vector, $M_{\sigma,i} \in \{0,1\}^{d_i \times d_i}$ is the transition matrix of $\sigma\in \Sigma$ and $\gamma_i \in \{0,1\}^{d_i \times 1}$ is the final column vector.
By construction, the weighted automaton $(\alpha_i,(M_{\sigma,i})_{\sigma\in \Sigma},\gamma_i)$ counts the number of runs of $D_i$ on a given word. Since each 
 $D_i$ is deterministic, the value of the weighted automaton on a word $w\in  \Sigma^*$ is 1 if $D_i$ accepts~$w$ and 0 otherwise.

The reduction constructs $3+|\Sigma|$ matrices of dimension $d:=2+\sum_{i=1}^n d_i$, namely $M_\mathrm{init},M_\mathrm{end}, M_t$, and $M_{\sigma}$ for each $\sigma\in \Sigma$. 
The matrix $M_t$ is the target matrix, and the question is to check whether $M_t$ lies in the monoid generated by the other matrices. 
Below, for simplicity of notation, we extend the map $\sigma \mapsto M_{\sigma,i}$ to a monoid homomorphism $\Sigma^{*}$ in a natural way (where for the empty word $\varepsilon$ the matrix $M_{\varepsilon,i}$ is the identity matrix for appropriate size).  Analogously, we extend the maps $\sigma \mapsto M_{\sigma}$ to $\Sigma^{*}$. 
The idea is to simulate the  computation of all weighted automata $(\alpha_i,(M_{\sigma,i})_{\sigma\in \Sigma},\gamma_i)$ over words $w$,
that is by definition~$\alpha_i M_{w,i} \gamma_i$, simultaneously by the matrix product
$M_\mathrm{init} M_w M_\mathrm{end}$.

To this end, we define 
the matrix $M_\mathrm{init}$ which  initializes all weighted automata simultaneously, and similarly
the matrix $M_\mathrm{end}$  which accounts for the final column vectors:

\[M_{\mathrm{init}} :=
\left(
\begin{array}{cccccc}
0 & \alpha_1 & \alpha_2 & \cdots & \alpha_n & 0 \\
\hline
 \multicolumn{6}{c}{\boldsymbol{0}_{(d-1)\times d}}
\end{array}
\right) \qquad \text{and} \qquad M_{\mathrm{end}} :=
\left(
\begin{array}{c|c}
\multirow{6}{*}{$\boldsymbol{0}_{d\times (d-1)}$} & 0 \\
 &
\gamma_1  \\
  & \gamma_2  \\
  & \vdots \\ 
 & \gamma_n\\
& 0 \\  
\end{array}
\right)
\]
For $\sigma\in \Sigma$, the matrix $M_{\sigma}$  simulates simultaneously applying the transition matrices $M_{\sigma,i}$,  relying on the block multiplication: 
\[
M_{\sigma} :=
\left(
\begin{array}{c|cccc|c}
 \multicolumn{6}{c}{\boldsymbol{0}_{1\times d}}
\\ \hline
\multirow{5}{*}{$\boldsymbol{0}_{d\times 1}$}
& M_{\sigma,1}
& \boldsymbol{0}_{d_1\times d_2}
& \cdots
& \boldsymbol{0}_{d_1\times d_n}
& \multirow{5}{*}{$\boldsymbol{0}_{d\times 1}$}
\\
& \boldsymbol{0}_{d_2\times d_1}
& M_{\sigma,2}
& \cdots
& \boldsymbol{0}_{d_2\times d_n}
& 
\\
&
\multicolumn{4}{c|}{\vdots}
&
\\
& \boldsymbol{0}_{d_n\times d_1}
& \boldsymbol{0}_{d_n\times d_2}
& \cdots
& M_{\sigma,n}
& 
\\ \hline
 \multicolumn{6}{c}{\boldsymbol{0}_{1\times d}}
\end{array}
\right)
\]
The target matrix $M_t$ is zero everywhere except in its top-right entry, where it is $n$.

We argue the correctness of the reduction as follows. Let \(X:=\{M_\mathrm{init},M_\mathrm{end}\} \cup \{M_{\sigma} \mid \sigma \in \Sigma\}\).
By the block matrix multiplication, \(X^{+} M_{\mathrm{init}} X^{\ast}=\{0\}\) and \(X^{\ast} M_{\mathrm{end}} X^{+}=\{0\}\). Hence we can restrict ourselves to products $M_{\mathrm{init}}^{\leq 1}M_{w} M_{\mathrm{end}}^{\leq 1}$ with $w\in \Sigma^{*}$. Moreover, we can assume the form \(M_{\mathrm{init}} M_w M_{\mathrm{end}}\), otherwise the top-right entry is \(0\) instead of \(n\).

By a simple induction, we can see that 
$M_{\mathrm{init}}M_{w}$, with $w\in \Sigma^{*}$, is in the following form:
\begin{equation}
\label{eq-weight}
    M_{\mathrm{init}} M_{w}:=
\left(
\begin{array}{ccccc}
0 & \alpha_1 M_{w,1}&  \cdots & \alpha_n M_{w,n}& 0 \\
\hline
 \multicolumn{5}{c}{\boldsymbol{0}_{(d-1)\times d}}
\end{array}
\right)
\end{equation}
where the base of induction ($w=\varepsilon$) holds by the definition of $M_{\mathrm{init}}$, and the induction step follows by block multiplication.
Given the form in~\eqref{eq-weight}, it follows that 
$M_{\mathrm{init}} M_{w} M_{\mathrm{end}}$ is a zero matrix, except in its top-right entry, where it stores $\sum_{i=1}^n \alpha_i M_{w,i} \gamma_i$. The latter is $n$ if and only if $w$ is accepted by all $D_i$. 
To conclude the proof, we note that entries of matrices in the monoid range over $\{0,\ldots, n\}$, hence the monoid is finite.

\textbf{Extension to semirings}: The proof extends for matrices over
any semiring $R\neq \{0\}$ and irreducible monoids as follows. To
extend to a semiring \(R \neq \{0\}\), we have to prevent addition
\(1+1\) from occurring, as its value would depend on the semiring. To
this end, we adapt the above construction as follows: We set
\(d:=1+n+\sum_{i=1}^n d_i\), where the extra dimensions are used as
additional target states. The idea is that instead of accumulating all
\(n\) runs on the same target state, for every DFA \(D_i\) we have a
separate target state which checks whether the corresponding DFA
accepts. Formally, we still use \(3+|\Sigma|\) matrices
\(M_{\mathrm{init}}, M_{\mathrm{end}}, M_{\mathrm{t}}\) and
\(M_\sigma\) for every \(\sigma\). The matrices \(M_{\mathrm{init}}\)
and \(M_{\sigma}\) are as defined above, extended with \(n-1\)
additional zero columns at the right. The matrices \(M_{\mathrm{t}}\)
and \(M_{\mathrm{end}}\) are now defined as follows, the left matrix
being \(M_{\mathrm{t}}\) and the right matrix is \(M_{\mathrm{end}}\).

\begin{align*}
&\left(
\begin{array}{c|c}
\multirow{2}{*}{$\boldsymbol{0}_{d\times (d-n)}$}
& \boldsymbol{1}_{1 \times n} \\
& \boldsymbol{0}_{(d-1) \times n}
\end{array}
\right) 
&&\left(
\begin{array}{c|ccccc}
\multirow{6}{*}{$\boldsymbol{0}_{(d-n)\times (d-n)}$} & 0 & 0 & \dots & 0 \\
 & \gamma_1 & 0_{d_1 \times 1} & \dots & 0_{d_1 \times 1} \\
 &0_{d_2 \times 1} & \gamma_2  &  & \vdots \\
& \vdots &  & \ddots & 0_{d_{n-1} \times 1}\\
&0_{d_n \times 1} & \dots & 0_{d_n \times 1} & \gamma_n\\  
\hline
\multicolumn{5}{c}{\boldsymbol{0}_{n\times d}}
\end{array}
\right)
\end{align*}

We argue correctness similarly to the above. Let
\(X:=\{M_\mathrm{init},M_\mathrm{end}\} \cup \{M_{\sigma} \mid \sigma
\in \Sigma\}\). Again, \(X^{+} M_{\mathrm{init}} X^{\ast}=\{0\}\) and
\(X^{\ast} M_{\mathrm{end}} X^{+}=\{0\}\). Hence we can restrict
ourselves to products $M_{\mathrm{init}}^{\leq 1}M_{w}
M_{\mathrm{end}}^{\leq 1}$ with $w\in \Sigma^{*}$. Again, we can
assume \(M_{\mathrm{init}} M_w M_{\mathrm{end}}\), as otherwise the
top-right entries are \(0\) instead of \(1\). With a similar induction
as before, we observe that $M_{\mathrm{init}} M_{w} M_{\mathrm{end}}$
is a zero matrix, except in the top-right entries, where the \(i\)-th
entry stores \(\alpha_i M_{w,i} \gamma_i\), i.e.\ is equal to \(1\) if
and only if the DFA \(D_i\) accepts \(w\). Hence all these entries are
equal to \(1\) as in \(M_{\mathrm{t}}\) if and only if \(w\) is accepted by all \(D_i\). Moreover, observe that every \(a \in X^{\ast}\) fulfills \(a \in \{0,1\}^{d \times d}\), because the only operations performed are \(0+0=0\), \(1+0=1\), \(1 \cdot 0=0\) and \(1 \cdot 1=1\) which hold in every semiring \(R\).

\textbf{Extension to Irreducible Finite Monoids}: A monoid
\(M \subseteq M_d(\mathbb{Q})\) is called \emph{irreducible} if there
is no vector space \(0 \subsetneq V \subsetneq \mathbb{Q}^d\) such
that \(V \cdot M \subseteq V\)\footnote{Defining irreducible with left
  or right multiplication gives the same definition, but right
  multiplication is more natural in automata theory}, i.e.\ no vector
space stable under multiplication by \(M\). We will show that
membership is \pspace-hard also for irreducible finite monoids \(M\).

\begin{proposition}
    Membership is \pspace-hard even under the promise that \(X^{\ast}\) is an \emph{irreducible} finite matrix monoid.
\end{proposition}

\begin{proof}
We again use \(d:=1+n+\sum_{i=1}^n d_i\) and base this construction on
the semiring reduction. As opposed to the above reductions though, we
now use \(3+|\Sigma|+2(d-1)\) many matrices. The first \(3+|\Sigma|\)
matrices are \(M_{\mathrm{init}}, M_{\mathrm{end}}, M_{\mathrm{t}}\)
and \(M_{\sigma}\) for every \(\sigma \in \Sigma\) as before, they
even have \emph{exactly} the same definition as in the semiring
reduction. Again, \(M_{\mathrm{t}}\) is the target matrix. The new
matrices are \(\{\delta_{1,j} \mid j \in \{2,\dots, d\}\}\) and
\(\{\delta_{j,1} \mid j \in \{2,\dots, d\}\}\), where \(\delta_{i,j}\)
is the Kronecker delta, i.e.\ the matrix with every entry \(0\) except
a \(1\) at the intersection of the \(i\)-th row and \(j\)-th column. These matrices are added purely to ensure irreducibility. Namely \(\delta_{j,1}\) resets the \(j\)-th state to the initial state, and is otherwise \(0\). The matrix \(\delta_{1,j}\) on the other hand is a type of initialization matrix which only initializes one DFA. None of these matrices will be useful for reaching the target, but they will ensure that no vector space can be stable under the monoid.

\textbf{Correctness}: First, observe that the monoid \(X^{\ast}\) is
still finite: In the semiring reduction we argued that
\(X^{\ast} \subseteq \{0,1\}^{d \times d}\), the same property can be
proven in the same way.

Next we observe that the \(\delta\)'s are not useful for reaching the target matrix. Namely, let \(w' \in X^{*_{concat}}\) be any word of generator matrices (we do not yet identify \(w'\) with an element of the monoid, hence we clarify that the \(\ast\) is concatenation here) which evaluates to \(M_{\mathrm{t}}\). We claim that there is a word \(w''\) without \(\delta\)'s which also evaluates to the target. 

Proof of claim: If \(w'\) does not contain \(\delta\)'s we are done. Hence assume that \(w'[k]=\delta_{i,j}\) for some \(i,j,k\), i.e.\ that some matrix used is \(\delta_{i,j}\). Then the product \(w'[1] \cdot \dots \cdot w'[k]\), i.e.\ the product corresponding to a prefix of the word, has only one non-zero column. However, the target matrix has multiple non-zero columns. By the structure of our matrices, in order to produce multiple non-zero columns from a single one, we must have \(w'[k']=M_{\mathrm{init}}\) for some \(k'>k\). In order for \(M_{\mathrm{init}}\) to not lead to the \(0\) matrix, the non-zero column must have been the first column. The only generators \(x \in X\) where the first column is non-zero are \(\delta_{j,1}\) for some \(j\). Hence we have \(w'[k'-1]=\delta_{j,1}\) for some \(j\). But then the product \(w'[1] \cdot \dots \cdot w'[k'-1]\) is useless. Removing it from the product, we obtain a new word \(w''=w'[k'] \dots w'[|w'|]\) with fewer \(\delta\)'s. Repeating the process, we eventually arrive at a word \(w''\) without any \(\delta\)'s, which still evaluates to the target, proving the claim.

By the claim, correctness now follows from the semiring correctness argument.

Finally, we have to argue that the monoid \(X^{\ast}\) is
irreducible. Let \(0 \subsetneq V\) be a vector space such that
\(V \cdot X \subseteq V\). We have to show that
\(V=\mathbb{Q}^d\). Since \(0 \subsetneq V\), there exists some
\(0 \neq v \in V\). Let \(j\) be such that \(v[j] \neq 0\). By
applying the matrix \(\delta_{j,1}\), we reach the vector
\((v[j],0,\dots, 0) \in V\). Since \(V\) is a vector space, i.e.\
closed under scalar multiplication, we have \(e_1 \in V\) where
\(e_1 \in \mathbb{Q}^d\) is the first unit vector. By applying
\(\delta_{1,j}\) we obtain \(e_j \in V\) also for the other unit
vectors, i.e.\ \(j=2, \dots, d\), which finishes the proof that
\(V=\mathbb{Q}^d\).
\end{proof}

\section{Integrality}
\label{sec:integrality}

For a finite set $X\subseteq \mat_d(\Q)$,
 the decision version of the Integrality Problem asks whether  $X$ is conjugate to a set of integral matrices,
 while the function version computes a conjugation matrix $p\in \gl_d(\mathbb{Q})$ such that $p^{-1}Xp\subseteq  \mat_d(\Z)$.
 Observe that a set $X$ is conjugate to a set of integral matrices if and only if this holds for the set $X^*$. 

\begin{theorem} \label{thm:integrality-ptime}
   The decision and function versions of the Integrality Problem lie in \ptime.
\end{theorem}

To prove Theorem \ref{thm:integrality-ptime}, we first 
extend \cite[Proposition 2.3]{BabaiFinitenessGroups} from groups to semigroups  following a  similar argument.

\begin{proposition}
\label{prop:zmodule}
    Let $X \subseteq \mat_d(\Q)$ be finite, and consider the $\mathbb Z$-module 
    $L$ generated by the set of vectors \(X^{\ast} \cdot \Z^d\).
    Then the following are equivalent:
    \begin{enumerate}
    \item  $L$ is a free $\mathbb Z$-module of rank $d$;
    \item  there exists $p \in \gl_d(\Q)$ such that $p^{-1}Xp \subseteq \mat_d(\Z)$;
    \item there exists a common denominator \(q\), i.e.\ $q\in \Z\setminus \{0\}$ such that $q \cdot X^{\ast} \subseteq \mat_d(\Z)$.
    \end{enumerate}
\end{proposition}

 \begin{proof}
  We first show that Item {\bf 1} implies Item {\bf 2}.
 Let $v_1,\ldots,v_d \in \Q^d$ be a $\Z$-basis of~$L$ and define 
$p \in \gl_d(\Q)$ as the matrix whose $i$-th column is $v_i$, for $i\in \{1,\ldots,d\}$. 
 Given $a\in X$, since $aL \subseteq L$,  for all $i\in \{1,\ldots,d\}$ there exists \(b_i \in \Z^d\)
 such that $a v_i =p b_i$. Defining \(b\) as the matrix with the columns \(b_i\) we obtain $ap=pb$. Hence, $p^{-1}ap=b\in \mat_d(\Z)$.

We next show that Item {\bf 2} implies Item {\bf 3}. Assume that there exists  $p \in \gl_d(\Q)$ such that $p^{-1}Xp \subseteq \mat_d(\Z)$. 
Then $X^{\ast} \subseteq p \mat_d(\Z)p^{-1}$. Hence 
$X^{\ast} \subseteq q^2 \mat_d(\Z)$ where 
$q\in \Z\setminus \{0\}$ is chosen such that $qp$ and $qp^{-1}$ are integer matrices. This establishes~(3). 

Finally, we show that Item {\bf 3} implies~Item {\bf 1}. Let $q\in \Z\setminus \{0\}$ be such that $q X^{\ast} \subseteq \mat_d(\Z)$. Then
$\Z^n \subseteq L \subseteq \frac{1}{q} \Z^n$.  
Since $L$ is a submodule of $\frac{1}{q}\Z^n$, which is free of rank $n$, it follows that $L$ is free of rank at most $n$.
Furthermore, since \(L\) contains $\Z^n$ it has rank exactly $n$.
\end{proof}

The idea behind the algorithm for   integrality is to  determine whether a given number is a common denominator of \(X^{\ast}\), extending \cite[Lemma 3.1]{BabaiFinitenessGroups} from groups to semigroups.

\begin{lemma} \cite[Lemma
  3.1]{BabaiFinitenessGroups} \label{lem:DecideIntegralityGivenDenominatorBound}
  There is a \ptime algorithm which, given a finite set
  \(X \subseteq \mat_d(\mathbb{Q})\) and a number \(q \in \N\) (in
  binary), decides whether \(q\) is a common denominator of
  \(X^{\ast}\), and if so computes a conjugation matrix
  \(p \in \gl_d(\mathbb{Q})\) such that
  \(p^{-1}Xp \subseteq \mat_d(\mathbb{Z})\).
\end{lemma}

\begin{proof}
Let $X\subseteq \mat_d(\Q)$, let \(m:=\bitsize(X)\) and let \(q\) be a common denominator. Let $L:=\spn_\Z(X^{\ast}\cdot \Z^d)\subseteq \Q^d\) be the \(\Z\)-module generated by \(X^{\ast}\cdot \Z^d\).
As shown in the proof of Item {\bf 1} implies  Item {\bf 2} in Proposition~\ref{prop:zmodule}, the columns of the required base change $p$ are a basis of $L$.

Observe that $q\Z^d \subseteq qL \subseteq \Z^d$.  To compute $qL$,
we define a sequence of vectors $(v_k)$ inductively as follows.  
First we choose $v_1,\ldots,v_d$ to be any set of vectors spanning $q\Z^d$.
Assume that we have defined $v_1,\ldots,v_k$ for some $k\in\mathbb N$.
If there exists \(a \in X\) and $l \in \{1,\ldots,k\}$
such that $a v_l$ does not lie in the $\Z$-span of $\{v_1,\ldots,v_k\}$ then we define
$v_{k+1}:=a v_l$ (the choice of \(a\) and $v_l$ doesn't matter). Otherwise we terminate the construction.
This sequence terminates in at most $d(\log_2 q)+\frac{d}{2}m$ steps by~\cite[Prop. 3.1]{BCSW24}
and upon termination the 
resulting sequence of vectors spans $qL$.  Computing the vector $v_{k+1}$ from $\{v_1,\ldots,v_k\}$ involves 
solving linear equations over~$\Z$, which is well-known to be in polynomial time (by the Hermite Normal Form~\cite{KannanBachem}).  Having obtained a generating set 
of $qL$, we can compute a basis thereof using the Smith Normal Form in polynomial time.
\end{proof}

Observe that Lemma \ref{lem:DecideIntegralityGivenDenominatorBound} does not yet imply Theorem \ref{thm:integrality-ptime}; we require a bound on the common denominator \(q\) which we can write using polynomially many bits and input to the algorithm of Lemma \ref{lem:DecideIntegralityGivenDenominatorBound}. We proceed to prove such a bound.

\begin{proposition}
  \label{thm:integrality-char}
Let \(X \subseteq \mat_d(\Q)\) be a finite set of matrices in \tnf. Let \(q'\) be a common denominator for the set \(X\) of generators. Let \(B \subseteq X^{\ast}\) be a basis of the enveloping algebra \(\enve(X)\). Assume the monoid \(X^{\ast}\) has some common denominator \(q \in \N\), then also \(q:=\det(\tmatrix_{\BlockDiag(B)})^d \cdot (q')^{d-1}\) is a common denominator for \(X^{\ast}.\)
\end{proposition}

\begin{proof}[Proof of Proposition~\ref{thm:integrality-char}]
    Let $X \subseteq \mat_d(\Q)$ be a finite set of matrices in \tnf
    such that
    \(X^{\ast}\) has a common denominator. Just as in Proposition \ref{prop:heightbound-wmnf}, write $X = \{a_1+c_1,\ldots,a_m+c_m\}\subseteq X^*$, where each $a_i$ is block diagonal and each $c_i$ strictly upper block triangular. Write  $A:=\{a_1,\ldots,a_m\}$, $C:=\{c_1,\ldots,c_m\}$ and $D:=\{a_1,\ldots,a_m,c_1,\ldots c_m\}$. We first claim that \(\det(\tmatrix_{\BlockDiag(B)})\) is a common denominator for \(A^{\ast}\).
    
    Proof of claim: We first show that \(\mathrm{tr}(a) \in \mathbb{Z}\) for all \(a \in X^{\ast}\). Since \(X^{\ast}\) has a common denominator \(q\), there exists a conjugation matrix \(p \in \gl_d(\Q)\) such that \(p^{-1}X^{\ast}p \subseteq \mat_d(\Z)\) by Proposition \ref{prop:zmodule}. In particular, \(p^{-1}ap \in \mat_d(\Z)\). Therefore \(tr(p^{-1}ap) \in \Z\), and since the trace is invariant under base change, we obtain \(tr(a) \in \Z\). This in particular implies that \(\tmap_B(a) \in \mathbb{Z}^{|B|}\) for all \(a \in X^{\ast}\).

    By definition of \tnf the block diagonal part \(a\) for \(a+c \in X^{\ast}\) can be uniquely determined from \(\tmap_B(a)\). Since \(\tmap_B(a) \in \Z^{|B|}\), the denominator of \(a\) hence divides the denominator of \(\tmap_B^{-1}\). By Cramer's rule, since \(\tmatrix_{\BlockDiag(B)}\) is an integer matrix, \(\det(\tmatrix_{\BlockDiag(B)})\) is a common denominator for \(\tmatrix_{\BlockDiag(B)}^{-1}\), finishing the claim.

     Since any matrix in $(A^*C)^d$ is zero, we obtain that \(q=(\det(\tmatrix_{\BlockDiag(B)}))^d \cdot (q')^{d-1}\) is a common denominator for \(D^{\ast}\). Since any matrix in \(X^{\ast}\) is by the distributive law a sum of matrices in \(D^{\ast}\), we obtain that \(q\) is also a common denominator for \(X^{\ast}\). 
\end{proof}

We can now prove Theorem \ref{thm:integrality-ptime}.
\begin{proof}[Proof of Theorem \ref{thm:integrality-ptime}]
By Theorem \ref{TheoremTraceNormalFormComputation} we can transform \(X^{\ast}\) to \tnf with polynomial overhead. Henceforth assume that \(X^{\ast}\) is in \tnf. 
Compute the number \(q=\det(\tmatrix_{\BlockDiag(B)})^d \cdot (q')^{d-1}\) of Proposition \ref{thm:integrality-char}, and output the result of applying Lemma \ref{lem:DecideIntegralityGivenDenominatorBound} with \((X,q)\) as input.
\end{proof}

\section{Discussion}
While it has been known for over 30 years that determining finiteness of finitely generated matrix groups is in polynomial time, until~\cite{BumpusHKST20} no elementary bound was known for the case of matrix monoids.
The present paper brings the complexity down to polynomial space.  It is an interesting question whether this can be further improved.  A more fine-grained optimisation of our results would be to improve the exponent in the polynomial bitsize bound in Theorem~\ref{thm:poly-bitsize}.
Another direction for further work concerns the complexity
of computing the Zariski closure of a finitely generated matrix monoid.  This is a generalisation of the problem of determining finiteness that has various applications in automata theory and
program analysis~\cite{HrushovskiOP018}.  Currently no elementary upper bounds are known for this problem.

\bibliography{bib-file}

\end{document}